\documentclass[12pt]{article}

\usepackage{amssymb}
\usepackage{graphicx}
\usepackage{amsmath}
\usepackage{amsfonts}
\usepackage[latin1]{inputenc}

\begin{document}

\begin{centering}
{\leftskip=2in \rightskip=2in
{\large \bf
Planck-scale structure of spacetime and some implications
for astrophysics and cosmology\footnote{Invited talk at ``Thinking,
Observing and Mining the Universe", Sorrento, Italy,
September 22-27, 2003} }}\\
\bigskip
\bigskip
\bigskip
\medskip
{\small {\bf Giovanni AMELINO-CAMELIA}}\\
\bigskip
{\it Dipart.~Fisica Univ.~La Sapienza and Sez.~Roma1 INFN}\\
{\it P.le Moro 2, I-00185 Roma, Italy}

\end{centering}

\vspace{0.7cm}

\begin{center}
\textbf{ABSTRACT}
\end{center}

\baselineskip 11pt plus .5pt minus .5pt

{\leftskip=0.6in \rightskip=0.6in {\footnotesize I briefly review
some scenarios for
the role of the Planck length in quantum gravity.
In particular, I examine the differences between
the schemes in which quantum
gravity is expected to introduce a maximum acceleration
and the schemes in which the Planck length sets the
minimum value of wavelengths (maximum value of momentum).
I also comment on some
pictures for the structure of spacetime at the Planck scale,
such as spacetime discretization and spacetime noncommutativity.
I stress that some of these proposals can have significant
implications in astrophysics and cosmology. }}

\newpage 
\baselineskip 12pt plus .5pt minus .5pt
\pagenumbering{arabic} 

\setcounter{footnote}{0} \renewcommand{\thefootnote}{\alph{footnote}}

\pagestyle{plain}

\section{The Planck length as a relativistic invariant}
One of the few (perhaps the only) rather robust hint that we have
about the quantum-gravity problem is that the
Planck length, $L_p \equiv \sqrt{\hbar G/c^3} \sim 10^{-35}m$,
should acquire a special role, and indeed
in most quantum-gravity research programmes
one finds or assumes that some new phenomena involve
in one or another way the Planck length
(or some closely-related scale, like the string length).
In recent years there has been growing interest (see, {\it e.g.},
Refs.~\cite{grbgac,gampul,billetal,kifu,gacmaj,polonpap,mexweave,ita,aus,gactp,jaco,gacpion,gacdsr,dsrnext,leedsr,dsrcosmo,thiemLS,seth,simonecarlo,areanew,schuacc,kodadsr,jurekkodadsr})
in establishing in which way the different scenarios for the
role of the Planck length in fundamental physics affect issues
that are relevant for rotation/boost transformations.
In particular, there are some cases in which the fact that
ordinary Lorentz boosts act nontrivially on lengths (FitzGerald-Lorentz
contraction) has been found to have profound consequences.

In these notes I intend to give a general discussion of the fate
of Lorentz symmetry in some illustrative examples of scenarios
for the role of the Planck length in fundamental physics.
And I will also briefly review some known cases in which observations
in astrophysics and cosmology are being considered as possible
opportunities for establishing experimentally the fate
of Lorentz symmetry at the Planck scale.

My first task must be the one of describing some
illustrative examples of scenarios
for the role of the Planck length in fundamental physics,
focusing on the implications for boost transformations.

\subsection{The Planck length as a coupling constant}
Of course, the Planck length already has a role
in the present (pre-quantum-gravity) description
of fundamental physics: it essentially
plays the role of a coupling constant (an $\hbar,c$-rescaled square root
of the gravitational coupling $G$).
Such a role for the Planck length is of course fully compatible
with the structure of ordinary Lorentz transformations.
Actually, a coupling constant can be given operative meaning
directly in a rest frame, and therefore, when it is operatively
defined in this way, it is fully compatible with all possible laws
of transformation between inertial observers.

I can clarify what I mean by ``can be given operative meaning
directly in a rest frame" by considering just the case of the
gravitational coupling constant: I can give operative meaning
to $G$ (and therefore to $L_p$) through a measurement of the force
that two massive particles at rest exert on each other.
Any observer that wants to establish the value of $G$ must consider
two particles at rest in her reference frame.
And of course, when operatively defined in this way, there is
clearly no logical obstruction for the physical law that ``the value
of $G$ is the same for all inertial observers".
In particular, this operative definition of $G$ (and $L_p$) is
fully (and trivially)
consistent with the structure of ordinary Lorentz transformations.

While it is legitimate to conjecture
that some new role for the Planck length should be introduced
at the level of quantum gravity,
it remains plausible that in quantum gravity,
just like in pre-quantum-gravity physics,
the role of the Planck length would be
simply the one of a coupling constant.
In particular, in String Theory the string length (which is closely
related to the Planck length) is introduced as a coupling constant
and appears to have
some rather familiar properties of coupling constants.

\subsection{The Planck length in a maximum-acceleration framework}
An example of new role for the Planck length is provided by scenarios
in which the Planck length intervenes in introducing a new
maximum-acceleration limit. To my knowledge the first discussions
of this type of scenario emerged in a research programme developed
by Caianiello and others~\cite{caia}. Actually the Caianiello programme
started off without invoking a role for the Planck length,
but rather seeking the introduction of a ``non-universal"
maximum-acceleration principle, with a different value of the maximum
acceleration for different particles. Since the speed-of-light scale
already provides us a velocity scale, the introduction of an acceleration
scale only requires a time(/length/inverse-mass) scale, a scale that
can be expressed in terms of the mass of the particle.
Eventually the Caianiello programme with
mass-dependent non-universal maximum acceleration encountered some
difficulties, and the Planck length was then used to introduce
a mass-independent universal maximum-acceleration limit.

More recently the possibility of a Planck-length-based
maximum-acceleration limit has
been explored in a new way~\cite{schuacc}, mostly on the basis
of the fact that in Born-Infeld theory there is a maximum field strength.

Of course, an observer-independent maximum-acceleration limit
can be introduced in a way that is fully compatible with ordinary
Lorentz transformations\footnote{But of course some nontrivial features must
be introduced for the description of accelerated observers (at
the General-Relativity level of analysis).}, since the acceleration of a particle
does not change\footnote{Of course, here the key point is that
acceleration are invariant under Lorentz transformations. Similar
schemes for introducing a length scale without modifying Lorentz
symmetry can be based on other Lorentz-transformation invariants
(see, {\it e.g.}, Ref.~\cite{kempscalar}).} under rotations and boosts.

This is somewhat analogous to the introduction of a minimum value
of the modulus of the angular-momentum vector
for some particles in ordinary quantum mechanics.
Such a minimum angular-momentum-modulus condition can be implemented,
as well known (and recently reconsidered from different perspectives
in Ref.~\cite{simonecarlo} and Ref.~\cite{areanew}),
without encountering any conflict with space-rotation symmetry.
Again this is due to the fact that
the modulus of the angular momentum of a particle does not
change under space rotations.

\subsection{The Planck length as the minimum wavelength}
While a role for the Planck length in the introduction of
a coupling constant or a maximum acceleration does not
require any modification of the familiar Lorentz transformations,
in some cases the Planck length can be introduced in a way
that requires modification of Lorentz boosts.
A good example of this possibility is provided by the
idea that the Planck length might set a
minimum-wavelength\footnote{The possibility
that the Planck length might set the minimum allowed value
of wavelengths has been considered in a large number of quantum-gravity
studies, but it is usually presented (see, {\it e.g.},
Refs.~\cite{kempmang,dadebro})
without commenting on the possibility that there might be
FitzGerald-Lorentz contraction of the minimum wavelength,
and without specifying how the minimum-wavelength scheme
is realized for different observers, {\it i.e.} without considering
the issue of whether the minimum value of wavelengths is the same
for all observers or it is given by the
Planck length only for one class of observers (while being
subject to FitzGerald-Lorentz contraction for other observers).}
(and/or maximum momentum) limit.

Since ordinary boosts act nontrivially on wavelengths (by FitzGerald-Lorentz
contraction) it is not possible
to introduce the Planck length as
an observer-independent minimum allowed value for wavelengths
without modifying the structure of Special Relativity.
Lorentz symmetry must be either ``broken" or ``deformed"
as one can easily see in a specific example of minimum-wavelength
scenario, the one in which the $\omega / \lambda$
frequency/wavelength dispersion relation (for a massless particle)
takes the form
\begin{equation}
\frac{L_p}{\lambda} = 1-e^{- L_p \omega/c}
~,
\label{disprejo}
\end{equation}
The possible implications of (\ref{disprejo}) for Special Relativity
can be easily analyzed in analogy with the possible implications
of the relation $E = \sqrt{c^2 p^2 + c^4 m^2}$ for Galilei-Newton Relativity.
According to Galilei-Newton Relativity the relativistic-invariant
relation between energy and momentum is $E=p^2/(2m)$, and therefore
there are only two possibilities:
if the Galilei-Newton rotation/boost transformations are not modified
the relation $E = \sqrt{c^2 p^2 + c^4 m^2}$ can only hold for
a single (``preferred") class of inertial observers (the ``ether");
if instead the relation $E = \sqrt{c^2 p^2 + c^4 m^2}$ is introduced
as an observer-independent law then the
Galilei-Newton rotation/boost transformations must necessarily be ``deformed".
The Lorentz rotation/boost transformations are a ``deformation"
of the Galilei-Newton rotation/boost transformations: they both are
6-parameter families of transformations, and in the $c^{-1} \rightarrow 0$
limit the Lorentz transformations are identical to the
Galilei-Newton transformations, but in general the Lorentz prescriptions
for relating observations in different reference frames differs from
the corresponding Galilei-Newton prescriptions.

Analogously for what concerns (\ref{disprejo}) there
are only two possibilities:
if the Lorentz rotation/boost transformations are not modified
the relation (\ref{disprejo}) can only hold for
a single (``preferred") class of inertial observers;
if instead the relation (\ref{disprejo}) is introduced
as an observer-independent law then the Lorentz
boost transformations must {\underline{necessarily}} be ``deformed".
The deformed Lorentz transformations would then be a
6-parameter family of symmetry transformations,
which in the $L_p \rightarrow 0$
limit are identical to the Lorentz transformations.

For the possibility of broken Lorentz symmetry (with a ``preferred"
class of inertial observers) there is a long tradition in the
quantum-gravity literature (see, {\it e.g.},
Refs.~\cite{grbgac,gampul,mexweave,ita,gactp,jaco}
and references therein).
The idea of a Planck-scale deformed Lorentz symmetry, in the
sense here described (in which the boost transformations
are characterized by two
observer-independent scales $c$ and $L_p$, rather
than the single invariant $c$), is the core feature of the
more recent ``Doubly Special Relativity"
proposal~\cite{gacdsr,dsrnext,leedsr}.

\subsection{The Planck length in ``quantum-gravity uncertainty principles"}
I have considered two rather different illustrative examples of situations
in which the Planck scale is introduced in ways that are completely
(and self-evidently) unconsequential for the analysis of Lorentz boosts,
the case of the Planck scale as a coupling constant and the case
of the Planck scale appearing in a maximum-acceleration limit.
I also considered an
illustrative example of a role for the Planck length that
necessarily requires some departures from Lorentz symmetry,
the case of the Planck length setting the minimum
allowed value for wavelengths (which requires that Lorentz symmetry
is either broken or deformed).

It is rather awkward that often in the quantum-gravity literature
some novel role is attributed to the
Planck length, without even commenting on the possibility
of FitzGerald-Lorentz contraction, without perceiving the need to
specify how the proposed new property
is realized for different (at least inertial) observers.
A very significant example of this incomprehensible practice
is provided by part of the literature on a possible role for the
Planck length in a new uncertainty principle for the measurement
of lengths. It is for example often
argued~\cite{mead,padma,dopl1994,ahlu1994,ng1994,gacmpla,garay,casadio}
that there should
be an uncertainty principle $min(\delta L) = L_p$ for the measurement of
any length $L$, and in most cases these proposals are discussed
without any comments on how this uncertainty principle should
be described by different observers. The key point here is that
there are important differences between the idea of
a $min(\delta L) = L_p$ limit for the measurement of proper lengths
(the length of, say, a pencil in its rest frame)
and the idea of a $min(\delta L) = L_p$
limit for the measurement of all lengths, but often the
proposals are formulated in the literature without even commenting
on which of these two very different scenarios is being pursued.

The introduction of a $min(\delta L) = L_p$
limit for the measurement of proper lengths
is of course inconsequential for Lorentz transformations.
It is a statement that acquires operative meaning
in a rest frame, just like the type of concept of coupling constant
that I considered earlier. If only the measurement of proper lengths
is affected by this minimum uncertainty it is then legitimate
to assume that other observers, boosted with respect to
the rest frame, will find a correspondingly smaller, FitzGerald-Lorentz
contracted, uncertainty.

If instead a relation of the type $min(\delta L) = L_p$ should
apply to all lengths (independently of the state of motion
of the object), one should then
contemplate the need for a deformation of the Lorentz
transformations, in the sense of the mentioned ``Doubly Special
Relativity" idea~\cite{gacdsr,dsrnext,leedsr}.

\section{The Planck length in discrete or noncommutative spacetimes}
FitzGerald-Lorentz contraction acts on physical lengths, but a
scale with dimensions of length can of course be introduced
in a way that its physical role is not the one of a length.
This is seen for example clearly in the case
of a theoretical framework with both a maximum velocity $V_{max}$
and a maximum acceleration $A_{max}$: from $V_{max}$
and $A_{max}$ one obtains the length scale $V_{max}^2/A_{max}$,
which (has clarified in the previous section) is not subject to
FitzGerald-Lorentz contraction. On the contrary, if, in an appropriate sense,
the Planck length is introduced at the level of the description
of the fundamental structure of spacetime it is instead natural
to expect nontrivial implications for Lorentz boosts.
Examples of roles for the Planck length in spacetime structure
are Planck-scale spacetime discreteness and
 Planck-scale spacetime noncommutativity.

 It is of course rather natural to explore the possibility
that in a quantum gravity some spacetime observables
be subject to noncommutativity and/or discretization,
since the solution of the quantum-gravity problem might require
the introduction of some characteristic features of quantum
theory in the description of spacetime.
Most types of
spacetime discretizations would be clearly incompatible
with the presence of an exact continuous (Lorentz) symmetry.
This is for example certainly the case~\cite{thooftdiscrete}
in approaches based on replacing the spacetime continuum
with a Planck-scale-discrete network of spacetime points.
But on the other hand it is clearly not true~\cite{simonecarlo,areanew}
that by introducing any element of discretization in a space one must
necessarily renounce to the presence of continuous symmetries.
For example the type of discretization of angular momentum
that is predicted by ordinary
quantum mechanics is fully consistent~\cite{areanew} with
invariance under space rotations.
It is therefore not possible to assume {\it a priori}
that any scenario for spacetime discretization considered
in the quantum-gravity
literature should lead to departures from Lorentz symmetry.

In particular over these past few years there has been intense
investigation of the fate of Lorentz symmetry in (the flat-spacetime
limit of) Loop Quantum Gravity, which (as presently understood)
predicts an inherently discretized spacetime~\cite{discreteareaLGQ}.
Although this discretization is not simply a description of spacetime
in terms of a discrete network of spacetime points (and therefore departures
from Lorentz symmetry are possible, but not automatically present),
arguments presented in Refs.~\cite{gampul,mexweave,thiemLS}
support the idea of broken Lorentz symmetry in Loop Quantum Gravity.
But the issue is not yet fully settled. In particular,
Ref.~\cite{simonecarlo} presents arguments in favour of unmodified
Lorentz symmetry in Loop Quantum Gravity, whereas
recently Smolin, Starodubtsev and I proposed~\cite{kodadsr}
(also see the related study in Ref.~\cite{jurekkodadsr})
a mechanism such that Loop Quantum Gravity
would be described at the most fundamental level as a theory that in the
flat-spacetime limit admits deformed Lorentz symmetry
(in the sense of Doubly Special Relativity).

Just like there are many ways in which one can introduce
some element of discretization in spacetime structure,
also spacetime noncommutativity can take many different forms.
Most studies have focused on various parts of the
following $Q_{\mu \nu},C^\beta_{\mu \nu}$
parameter space ($\mu,\nu,\beta = 0,1,2,3$)
\begin{equation}
\left[x_\mu,x_\nu\right] = i L_p^2 Q_{\mu \nu}
+ i L_p C^\beta_{\mu \nu} x_\beta ~,
\label{alllp}
\end{equation}
where $Q$ and $C$ are dimensionless matrices.
It is at this point clear, in light of several recent results,
that the only way to preserve Lorentz symmetry
is the choice $Q = 0 =C $ ({\it i.e.} the case in which
there is no noncommutativity
and one is back to the familiar classical
commutative Minkowski spacetime).
When noncommutativity is present
Lorentz symmetry is usually
broken, but for some special choices of the
matrices $Q$ and $C$
Lorentz symmetry might be deformed, rather than broken.

We have a rather detailed
understanding~\cite{susskind,dineIRUV,gacluisa,carlson,dougnekr}
 of the way in which
Lorentz symmetry is broken in
the ``canonical noncommutative
spacetimes''~\cite{wessLANGUAGE} with $C^\beta_{\mu \nu} =0$
($\left[x_\mu,x_\nu\right] = i L_p^2 Q_{\mu \nu}$).
The matrix $Q_{\mu \nu}$ transforms like a tensor in going from
one inertial observer to another (the noncommutativity is
observer dependent). Particles progating in these
canonical spacetimes are governed by an energy/momentum
dispersion relation which is $Q$ dependent and different
from $E=\sqrt{c^2 p^2 + c^4 m^2}$.
And there are birefringence effects: different polarizations
of light travel at different speeds, just as it happens in
the study of the propagation of light in some material crystals.

An example of noncommutative spacetime in which Lorentz symmetry
might be deformed rather than broken
is $\kappa$-Minkowski~\cite{gacmaj,majrue,kpoinap,lukieFT,gacmich,wesskappa}:
\begin{equation}
\left[x_\mu, x_\nu \right] = i L_p ( \delta_{\mu}^\beta \delta_{\nu}^0
- \delta_{\nu}^\beta \delta_{\mu}^0 )
\label{kmindef}
\end{equation}
({\it i.e.} $[x_j, x_0] = i L_p x_j$, $[x_m, x_l] = 0$).
It appears\footnote{Most of the properties so far uncovered
for $\kappa$-Minkowski are consistent with the structure of the
mentioned ``Doubly Special Relativity"
framework~\cite{gacdsr,dsrnext,leedsr}.
There are however some open issues for the compatibility
of $\kappa$-Minkowski with Doubly Special Relativity,
mostly concerning systems of two or more particles~\cite{areanew}.}
possible to introduce this  $\kappa$-Minkowski
noncommutativity as an observer-independent description of
spacetime, at the price of a deformation of the Lorentz transformations.
Particles progating in $\kappa$-Minkowski
are governed by an energy/momentum
dispersion relation which is $L_p$ dependent (but observer independent)
and different from $E=\sqrt{c^2 p^2 + c^4 m^2}$.

\section{Some implications for astrophysics and cosmology}
Quantum-gravity effects are extremely small, since their magnitude
is typically set by some power of the ratio between the Planck scale
and the wavelength of the particle under study.
There are some contexts in which the theoretical predictions
can be confronted with data, but in most cases it is necessary
to rely on observations in astrophysics and cosmology, rather
than laboratory experiments~\cite{polonpap}.
The ``astrophysics of quantum gravity" is being considered
also for effects that are not directly related to the issues
for Lorentz symmetry that I discussed here (see, for example,
the Equivalence-Principle tests considered
in Refs.~\cite{gasperiniEP,dharamEP}),
and has been advocated in a large number of papers on the fate of
Lorentz symmetry in quantum gravity.
It is at this point well established that, if Lorentz symmetry is
broken or deformed at the Planck scale,
there are at least a handful of opportunities
for controntation with data.

One of the most studied scenarios is based on a modified
dispersion relation of the type (\ref{disprejo}),
using the associated small wavelength dependence of the speed of photons
(based on the relation $v = d\omega/dk$, $k \equiv 1/\lambda$).
The wavelength dependence of the speed of photons
that is induced by (\ref{disprejo}) is of order $L_p/ \lambda$,
and is therefore completely negligible in nearly all physical
contexts. It is however quite significant~\cite{grbgac,billetal}
for the analysis of short-duration gamma-ray bursts that reach
us from cosmological distances.
For a gamma-ray burst a typical estimate of the time travelled
before reaching our Earth detectors is $T \sim 10^{17} s$.
Microbursts within a burst can have very short duration,
as short as $10^{-4} s$.
Some of the photons in these bursts
have energies in the $100 MeV$ range and higher (and correspondingly small
wavelengths).
For two photons with energy difference of order $\Delta E \sim 100 MeV$
an $L_p \Delta E$ speed difference over a time of travel of $10^{17} s$
leads to a relative time-of-arrival delay of
order $\Delta t \sim \eta T L_p \Delta E \sim 10^{-3} s$.
Such a quantum-gravity-induced time-of-arrival delay
could be revealed~\cite{grbgac,billetal}
upon comparison of the structure of the gamma-ray-burst signal
in different energy channels, and these types of studies are planned
for the next generation of gamma-ray telescopes,
such as GLAST~\cite{glast}.

With advanced planned neutrino observatories, such as
ANTARES~\cite{antares}, NEMO~\cite{nemo} and EUSO~\cite{euso},
it should be possible to observe neutrinos with energies
between $10^{14}$ and $10^{19}$ $eV$,
and according to
current models~\cite{grbNEUTRINOnew}
gamma-ray bursters should also emit a substantial amount of
high-energy neutrinos. This might provide~\cite{grbNUnick,grf03ess}
another opportunity for time-of-arrival analyses.

Another example of opportunity to test schemes for Planck-scale
departures from Lorentz symmetry is the one of ``threshold
anomalies". For example, on the basis of a Planck-scale modified
energy-momentum dispersion relation that at low energies takes the
form $E^2 \simeq c^2 p^2 + c^4 m^2 + L_p E p^2$ (which could
be inspired by (\ref{disprejo})),
and assuming unmodified law of energy-momentum conservation
(which is compatible with the modified dispersion relation in
a scheme with Lorentz-symmetry breaking),
one finds that a collision between
a soft photon of energy $\epsilon$
and a high-energy photon of energy $E$ can produce
an electron positron pair ($\gamma + \gamma \rightarrow e^+ + e^-$)
only if $E \ge E_{th}$,
with the threshold energy $E_{th}$ given by~\cite{gactp}
\begin{equation}
E_{th} \epsilon - \eta L_p E_{th}^{3}/8 \simeq m_e^2
~.
\label{thrTRE}
\end{equation}
Analogous modifications of threshold relations are found for other
processes. In particular, the case of photopion
production, $p + \gamma \rightarrow p + \pi$,
also leads to an analogous result in the case in which the
incoming proton has high energy $E$ while the incoming
photon has energy $\epsilon$ such that $\epsilon \ll E$.
And the photopion-production threshold is relevant
for the analysis of UHE (ultra-high-energy) cosmic rays,
since a characteristic feature of the expected cosmic-ray spectrum,
the so-called ``GZK limit", depends on the evaluation of the
minimum energy required of a cosmic ray in order to produce pions
in collisions with cosmic-microwave-background photons.
Strong interest was generated by the
observation~\cite{kifu,ita,aus,gactp,jaco,alfaro,orfeupion}
that a Planck-scale-modified threshold relation can
lead to a
significantly higher estimate of the
threshold energy, resulting an upward shift of the GZK limit.
This would provide a description of the observations of the high-energy
cosmic-ray spectrum reported by AGASA~\cite{agasa},
which can be interpreted as an indication of
a sizeable upward shift of the GZK limit.
(But I must stress here that there are other plausible theory
explanations for the AGASA ``cosmic-ray puzzle",
and the experimental side must be further explored, since another
cosmic-ray observatory, HIRES, has not confirmed the AGASA results.)

An example of application of ideas for the fate of Lorentz at the
Planck scale in cosmology is given by the studies in Ref.~\cite{dsrcosmo}.
The observation that some of the frontier ideas in Planck-scale-physics
research can be important in cosmology
has already a rather long tradition (see, {\it e.g.},
Ref.~\cite{veneziano,huang,cremonini,casadiocosmo}).
In Ref.~\cite{dsrcosmo}
it is observed that the recent proposal of Doubly Special Relativity
schemes provides an opportunity for a reformulation of
the ``time-varying speed of light"
cosmological scenario~\cite{vsl}
(which had been previously structured relying on a preferred frame).


\end{document}